\pdfoutput=1
\RequirePackage[l2tabu,orthodox]{nag}
\documentclass[a4paper,11pt]{article}
\usepackage[T1]{fontenc}
\usepackage[utf8]{inputenc}
\usepackage{pos}
\usepackage{microtype}
\usepackage{mathtools}
\usepackage{csquotes}
\usepackage{physics}
\usepackage{slashed}
\usepackage{simpler-wick}
\usepackage{booktabs}
\usepackage{siunitx}

\usepackage[style=base]{caption}
\usepackage{subcaption}
\NewDocumentCommand\cpi{}{\uppi}                        
\NewDocumentCommand\I{}{\mathrm{i}}
\NewDocumentCommand\e{ m }{\mathrm{e}^{#1}}

\makeatletter
\DeclareFontFamily{OMX}{MnSymbolE}{}
\DeclareSymbolFont{MnLargeSymbols}{OMX}{MnSymbolE}{m}{n}
\SetSymbolFont{MnLargeSymbols}{bold}{OMX}{MnSymbolE}{b}{n}
\DeclareFontShape{OMX}{MnSymbolE}{m}{n}{
    <-6>  MnSymbolE5
   <6-7>  MnSymbolE6
   <7-8>  MnSymbolE7
   <8-9>  MnSymbolE8
   <9-10> MnSymbolE9
  <10-12> MnSymbolE10
  <12->   MnSymbolE12
}{}
\DeclareFontShape{OMX}{MnSymbolE}{b}{n}{
    <-6>  MnSymbolE-Bold5
   <6-7>  MnSymbolE-Bold6
   <7-8>  MnSymbolE-Bold7
   <8-9>  MnSymbolE-Bold8
   <9-10> MnSymbolE-Bold9
  <10-12> MnSymbolE-Bold10
  <12->   MnSymbolE-Bold12
}{}

\let\lAngle\@undefined
\let\rAngle\@undefined
\DeclareMathDelimiter{\lAngle}{\mathopen}%
                     {MnLargeSymbols}{'164}{MnLargeSymbols}{'164}
\DeclareMathDelimiter{\rAngle}{\mathclose}%
                     {MnLargeSymbols}{'171}{MnLargeSymbols}{'171}
\makeatother

\DeclareDocumentCommand\evsub{ s s m g }
{ 
	\IfNoValueTF{#4}
	{
		\IfBooleanTF{#1}
		{\vphantom{#3}\left\lAngle\smash{#3}\right\rAngle} 
		{\left\lAngle{#3}\right\rAngle} 
	}
	{
		\IfBooleanTF{#1}
		{
			\IfBooleanTF{#2}
			{\left\lAngle{#4}\middle\vert{#3}\middle\vert{#4}\right\rAngle} 
			{\vphantom{#3#4}\left\lAngle\smash{#4}\middle\vert\smash{#3}\middle\vert\smash{#4}\right\rAngle} 
		}
		{\vphantom{#3}\left\lAngle{#4}\middle\vert\smash{#3}\middle\vert{#4}\right\rAngle} 
	}
}

\title{Approaching the master-field:\\ Hadronic observables in large volumes}
\ShortTitle{Approaching the master-field: Hadronic observables in large volumes}

\author*[a]{Marco Cè}
\author[b,c]{Mattia Bruno}
\author[d]{John Bulava}
\author[e,a]{Anthony Francis}
\author[f]{Patrick Fritzsch}
\author[f,a]{Jeremy R.\ Green}
\author[g]{Maxwell T.\ Hansen}
\author[h,a]{Antonio Rago}

\affiliation[a]{Department of Theoretical Physics, CERN, 1211 Geneva 23, Switzerland}
\affiliation[b]{Dipartimento di Fisica \enquote{Giuseppe Occhialini}, Università degli Studi di Milano-Bicocca, Piazza della Scienza 3, 20126 Milan, Italy}
\affiliation[c]{INFN - Sezione di Milano Bicocca, Piazza della Scienza 3, 20126 Milan, Italy}
\affiliation[d]{Deutsches Elektronen-Synchrotron (DESY), Platanenallee 6, 15738 Zeuthen, Germany}
\affiliation[e]{Albert Einstein Center for Fundamental Physics and Institute for Theoretical Physics, Universität Bern, Sidlerstrasse 5, 3012 Bern, Switzerland}
\affiliation[f]{School of Mathematics and Hamilton Mathematics Institute, Trinity College Dublin, Dublin 2, Ireland}
\affiliation[g]{Higgs Centre for Theoretical Physics, School of Physics and Astronomy, The University of Edinburgh, Edinburgh EH9 3FD, United Kingdom}
\affiliation[h]{Centre for Mathematical Sciences, Plymouth University, Plymouth, PL4 8AA, United Kingdom}

\emailAdd{marco.ce@cern.ch}

\abstract{%
  The master-field approach to lattice QCD envisions performing calculations on a small number of large-volume gauge-field configurations.
  Substantial progress has been made recently in the generation of such fields, and this must be joined with measurement strategies that take advantage of the large volume.

  In these proceedings, we describe how to compute simple hadronic quantities efficiently and estimate their errors in the master-field approach, \emph{i.e.}\ by studying cross-correlations of observables on a single configuration.
  We discuss the scaling of the uncertainty with the volume and compare extractions based on momentum-projected and position-space two-point functions.
  The latter show promising results, already at intermediate volumes, but come with additional technical complexities such as a more complicated manifestation of boundary effects, which we also address.

  \begin{flushright}
    CERN-TH-2021-164
  \end{flushright}
}

\FullConference{%
 The 38th International Symposium on Lattice Field Theory, LATTICE2021
  26th-30th July, 2021
  Zoom/Gather@Massachusetts Institute of Technology
}

\begin{document}
\maketitle

\section{Introduction}

The term \emph{master-field} encapsulates the idea of estimating the expectation values of observables using the translation averages over a single (or a few) representative gauge field(s), the master-field(s)~\cite{Luscher:2017cjh}, instead of traditional estimators over an ensemble of gauge field configurations generated with Monte Carlo (MC) methods.
This relies on the \emph{stochastic locality} of the fields, that is, the fact that thanks to the mass gap, distant regions of a physically large lattice fluctuate largely independently.
Clearly, the realization of this idea requires lattices that have a size that is large compared with the finite correlation length.

Thanks to a number of algorithmic achievements \cite{Francis:2019muy}, combined with the increase of the available computational resources, simulations are currently underway for dynamical QCD lattices with volumes up to $(\SI{18}{\femto\metre})^4$ and $m_\pi L\approx\num{25}$, see Ref.~\cite{Fritzsch:2021lattice}.
This opens the door to actual practical computations based on translation averages.
However, while master-field computations of bosonic quantities in Yang-Mills theory have already being performed~\cite{Giusti:2018cmp}, the application to hadronic observables introduces additional challenges.
Here we present progress in solving the problem, reviewing the theory of master-field errors and studying the extraction of simple hadronic quantities, with a focus on position-space correlation functions and performing numerical tests on a traditional MC ensemble of lattice configurations with a large volume.

\section{Master-field errors}

In the master-field approach, the MC ensemble average of a local field $O(x)$ is replaced as an estimator of $\ev{O(x)}$ by the \emph{translation average}~\cite{Luscher:2017cjh}
\begin{equation}
  \evsub{O(x)} = \frac{1}{V} \sum_z O(x+z) , \qquad \ev{O(x)} = \evsub{O(x)} + \order{V^{-1/2}} .
\end{equation}
An error on this estimator can be derived by writing down the field-theoretical expression for the variance of its distribution, that is, the connected correlator of $O(x)$,
\begin{equation}
\begin{split}
\label{eq:mf_var_O}
  \sigma^2_{\evsub{O}}(x) &= \ev{ [\evsub{O(x)} -\ev{O(x)}]^2 }
                           = \frac{1}{V} \sum_{y} \ev{O(y) O(0)}_c \\
                          &= \frac{1}{V} \left[ \sum_{\abs{y}\leq R} \ev{O(y) O(0)}_c + \order{\e{-mR}} \right] \\
                          &= \frac{1}{V} \left[ \sum_{\abs{y}\leq R} \evsub{O(y) O(0)}_c + \order{\e{-mR}} + \order{V^{-1/2}} \right] ,
\end{split}
\end{equation}
where in the second line we used the fact that $O(x)$ is a local field and its connected correlator decays exponentially with spacetime separation, and in the last line we replaced the field-theoretical average with the master-field translation average introducing an error suppressed with the inverse square root of the volume.

If $n$ master-fields are available, a similar error formula is obtained for the translation average of the ensemble mean $\bar{O}(x)$,
\begin{equation}
  \sigma^2_{\evsub*{\bar{O}}}(x) = \frac{1}{n} \sigma^2_{\evsub{O}}(x) 
                                 = \frac{1}{V} \left[ \sum_{\abs{y}\leq R} \evsub{\bar{O}(y)\bar{O}(0)}_c + \order{\e{-mR}} + \order{V^{-1/2}} \right] .
\end{equation}

\subsection{Hadronic observables}

Simple hadronic quantities such as the masses of stable hadrons or meson decay constants, as well as more complex ones such as hadronic contributions to $g-2$ of the muon and real-time spectral functions, all generally require the computation of quark propagators from the solution of the Dirac equation.
For instance, consider the meson propagator with a single connected Wick contraction,
\begin{equation}
  C_{\Gamma\Gamma'}(x,0) = \wick{[\c1{\bar{u}}\Gamma\c2{d}](x) [\c2{\bar{d}}\Gamma' \c1{u}]}(0) = -\tr{ \gamma_5\Gamma\: D^{-1}(x,0)\: \Gamma'\gamma_5\: D^{-1}(x,0)^\dagger } .
\end{equation}
The quark propagator $D^{-1}(x,0)$ is non-zero at any $x$ in spacetime, so it is not an \emph{ultralocal} field but, empirically, on each representative gauge field its norm $\norm*{D^{-1}(x,0)}\propto\e{-m_\pi\abs{x}/2}$ for large $\abs{x}$, which localizes the quark propagator field to a region $\sim m_\pi^{-1}$.

Applying the master-field treatment to it, we have a translation-average estimator $\evsub{C(x,0)}$\footnote{
  We drop the subscript $\Gamma$ and $\Gamma'$ when referring to a generic Dirac structure.
}
whose error can be estimated according to Eq.~\eqref{eq:mf_var_O}
\begin{equation}
\label{eq:mf_var_C}
  \ev{ [\evsub{C(x,0)} -\ev{C(x,0)}]^2 } = \frac{1}{V} \left[ \sum_{\abs{y}\leq R} \evsub{C(x+y,y) C(x,0)}_c + \order{\e{-mR}} + \order{V^{-1/2}} \right] ,
\end{equation}
in terms of the translation average of the product $C(x+y,y) C(x,0)$.\footnote{%
  This \enquote{four-point} function can be obtained as the disconnected ($2+2$) Wick contraction of four meson fields.%
}

We note that, if all-to-all correlators are not available as it is generally the case, the translation averages and the truncated sum of the connected correlator in the error formula can still be computed sampling the source positions $y$.

\section{Position-space correlators}

Usually, momentum-projected correlators are used to extract specific hadronic observables. 
This is a viable strategy also with the master-field approach.
For instance, one can introduce the momentum projection at the source, \emph{e.g.}\ with a stochastic estimator, and define the momentum-projected correlator $\tilde{C}(x_0-y_0,\vec{p}) = \sum_{\vec{y}} \e{-\I\vec{p}\cdot(\vec{x}-\vec{y})}C(x, y)$.
The correlator obtained is a quasi-local function of $x$ at fixed $x_0-y_0$ and its error can be estimated using translation averages as in Eq.~\eqref{eq:mf_var_C}.

In these proceedings, however, we focus on the alternative option of exacting the hadronic masses directly from the the \emph{position-space} correlator $C(x)\equiv\ev{C(x,0)}$.
In the continuum and infinite volume, the two-point function in position-space of non-singlet pseudoscalar densities $P(x)$ tends to the bosonic single-particle correlator described by a modified Bessel function of the second kind,\footnote{%
  Or with $m_\pi$ replaced by $m_K$ depending on the flavour of $P(x)$.
  For simplicity, here we consider only the $m_\pi$ case.
}
\begin{equation}
\label{eq:CPPdef}
  C_{PP}(x) \to \frac{\abs{c_P}^2}{4\pi^2} \frac{m_\pi}{\abs{x}} K_1(m_\pi|x|) , \qquad\text{for $x\to\infty$} ,
\end{equation}
while the two-point function of nucleon interpolating fields $N(x)$ tends to
\begin{equation}
\label{eq:CNNdef}
  C_{NN}(x) \to \frac{\abs{c_N}^2}{4\pi^2} \frac{m_N^2}{\abs{x}}
                \left[ K_1(m_N\abs{x}) + \frac{\slashed{x}}{\abs{x}} K_2(m_N\abs{x}) \right] , \qquad\text{for $x\to\infty$} .
\end{equation}
The nucleon propagator in Eq.~\eqref{eq:CNNdef} is a Dirac spinor, and the different Dirac index contractions result in two different scalar functions for $x\to\infty$
\begin{subequations}
\begin{gather}
  \tr C_{NN}(x) \to \frac{\abs{c_N}^2}{4\pi^2} \frac{m_N^2}{\abs{x}} K_1(m_N\abs{x}) , \label{eq:trCNNdef} \\
  \tr\slashed{x}C_{NN}(x) \to \frac{\abs{c_N}^2}{4\pi^2} m_N^2 K_2(m_N\abs{x}) . \label{eq:trxCNNdef} 
\end{gather}
\end{subequations}

These equations give access to $m_\pi$ and $m_N$, but the rotationally-invariant descriptions are valid in the continuum limit, while at finite lattice spacing they are modified by contributions that depend on the direction of $x$.
We postpone a detailed study of the symmetry-breaking discretization effects and consider in the following only the correlator averaged over $4d$ spheres of given radius, introducing the correlatoion function of the radial coordinate $r$
\begin{equation}
  \mathring{C}(r) = \frac{1}{\mathrm{r}_4(r^2/a^2)} \sum_{\abs{x}=r} C(x) 
\end{equation}
for both $C_{PP}$ and $C_{NN}$, where $\mathrm{r}_4(n)=8\sum_{d\mid n,4\nmid d} d$ is the number of representations of $n\in\mathbb{N}$ as the sum of four squares, where representations that differ only in the order of the summands or in the signs of the numbers being squared are counted as different.\footnote{\url{https://oeis.org/A000118}}

\section{Numerical investigation}

We test the computation of $m_\pi$ and $m_N$ on an ensemble of \num{82} gauge field configurations of a lattice with tree-level Symanzik-improved Lüscher-Weisz gauge action and non-perturbatively $\order*{a}$-improved exponentiated-clover Wilson fermions~\cite{Francis:2019muy}, a spatial extent $L=64a$ and temporal extent $T=96a$, generated using the stochastic molecular dynamics (SMD) algorithm implemented in the \texttt{openQCD-2.0} software\footnote{\url{https://cern.ch/luscher/openQCD}}~\cite{Luscher:2012av}.
The bare gauge coupling $\beta=6/g_0^2=\num{3.8}$ corresponds to a lattice spacing $a\approx\SI{0.094}{\femto\metre}$, and the light and strange hopping parameters $\kappa_{\ell}=\num{0.1391874}$, $\kappa_s=\num{0.1385164}$ correspond to a pion and kaon mass of approximatively \SI{290}{\MeV} and \SI{450}{\MeV} respectively~\cite{Francis:2019muy}.

The spatial extent of the lattice is about \SI{6.08}{\femto\metre} in physical units, and $m_\pi L=\num{8.9}$ using the pion mass that we determine in Eq.~\eqref{eq:pion_result}.
The volume is too small to reliably apply the master-field error estimation, therefore the error on the numerical results obtained here is estimated from the traditional MC ensemble gauge variance.

To provide a realistic comparison to standard modern techniques for extracting hadron (especially baryon) masses, we apply smearing to the quark fields.
In particular, we implement $3d$-fermion smearing~\cite{Papinutto:2018ajw} with $\kappa_{3d}=\num{0.180}$, \num{0.190}, \num{0.200}, which works for momentum-projected correlators but breaks the four-dimensional hypercubic symmetry of position-space correlators.
In addition, we consider gradient-flow smearing of both source and sink quark fields, as defined in Ref.~\cite{Luscher:2013cpa}.
This preserves the symmetry of position-space correlators but it affects the transfer matrix and introduces effects that make the spectral representation of the pion and nucleon correlators non-positive.
However, these effects have a short range and we can ignore them as long as the smearing radius $\sqrt{8t_{\mathrm{flow}}}\approx\SI{0.3}{\femto\metre}$ is much smaller than the length scales at which the ground states start to dominate the correlator. 
We employ point sources and only for the time-momentum meson correlators also stochastic wall sources.

In the following, given a radial correlator we compute its effective mass $m_{\mathrm{eff}}(r)$ by numerically solving at every value of $r$ the equation
\begin{equation}
\label{eq:effective_mass_def}
   \eval{\frac{\mathring{C}(r+d; m)}{\mathring{C}(r; m)}}_{\mathrm{ansatz}} = \eval{\frac{\mathring{C}(r+d)}{\mathring{C}(r)}}_{\mathrm{data}}
\end{equation}
for the mass parameter $m$ in the appropriate correlator description.
The value of $d$ is chosen as the value closest to one unit of the lattice spacing $a$ such that $(r+d)^2/a^2\in\mathbb{N}$, as we observe that this produces an effective mass much smoother than using, for instance, the smallest possible value of $d$.

\subsection{Results for the pion mass}

\begin{figure}[t]
  \centering
  \resizebox{\columnwidth}{!}{\includegraphics{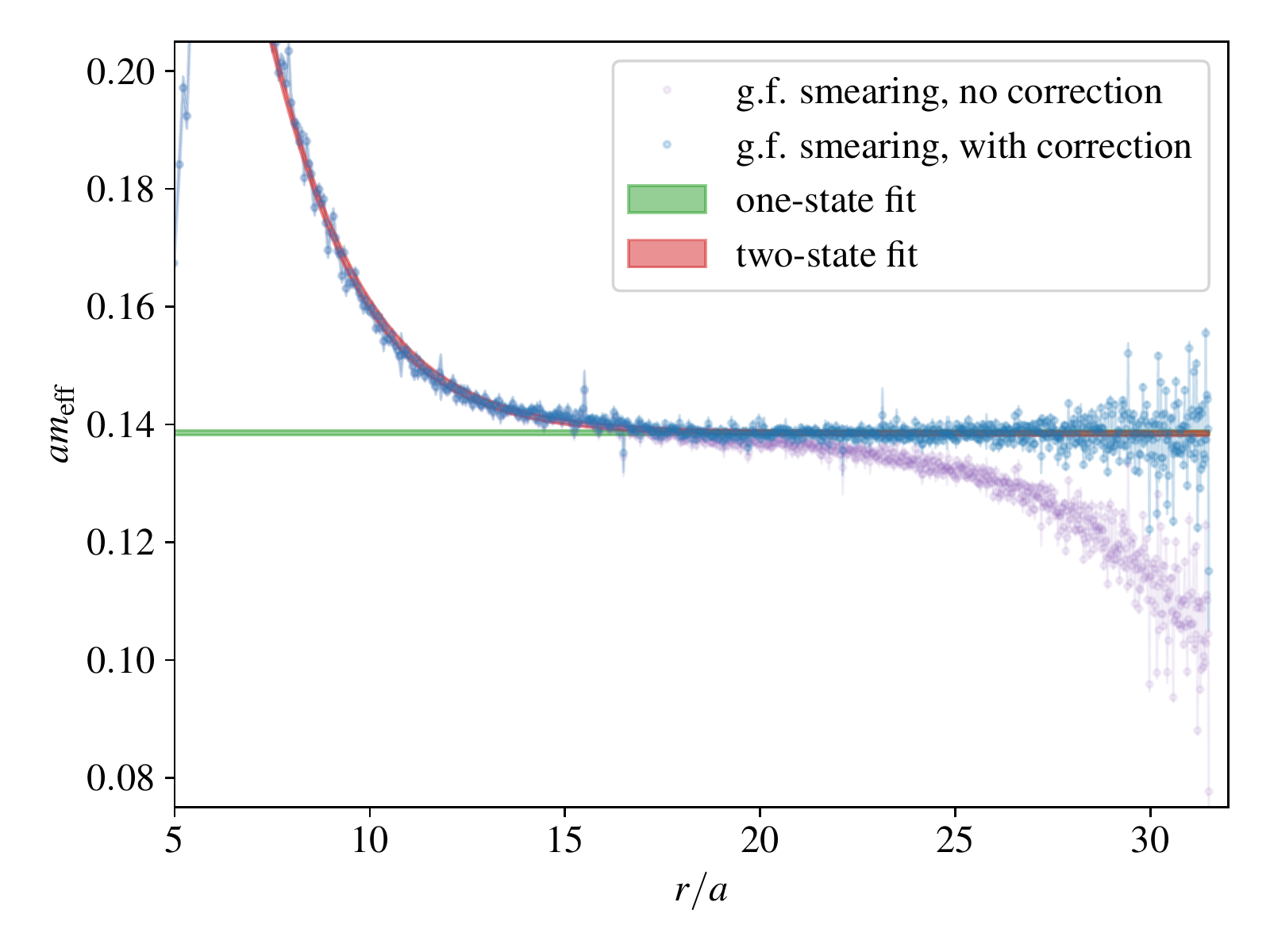}\includegraphics{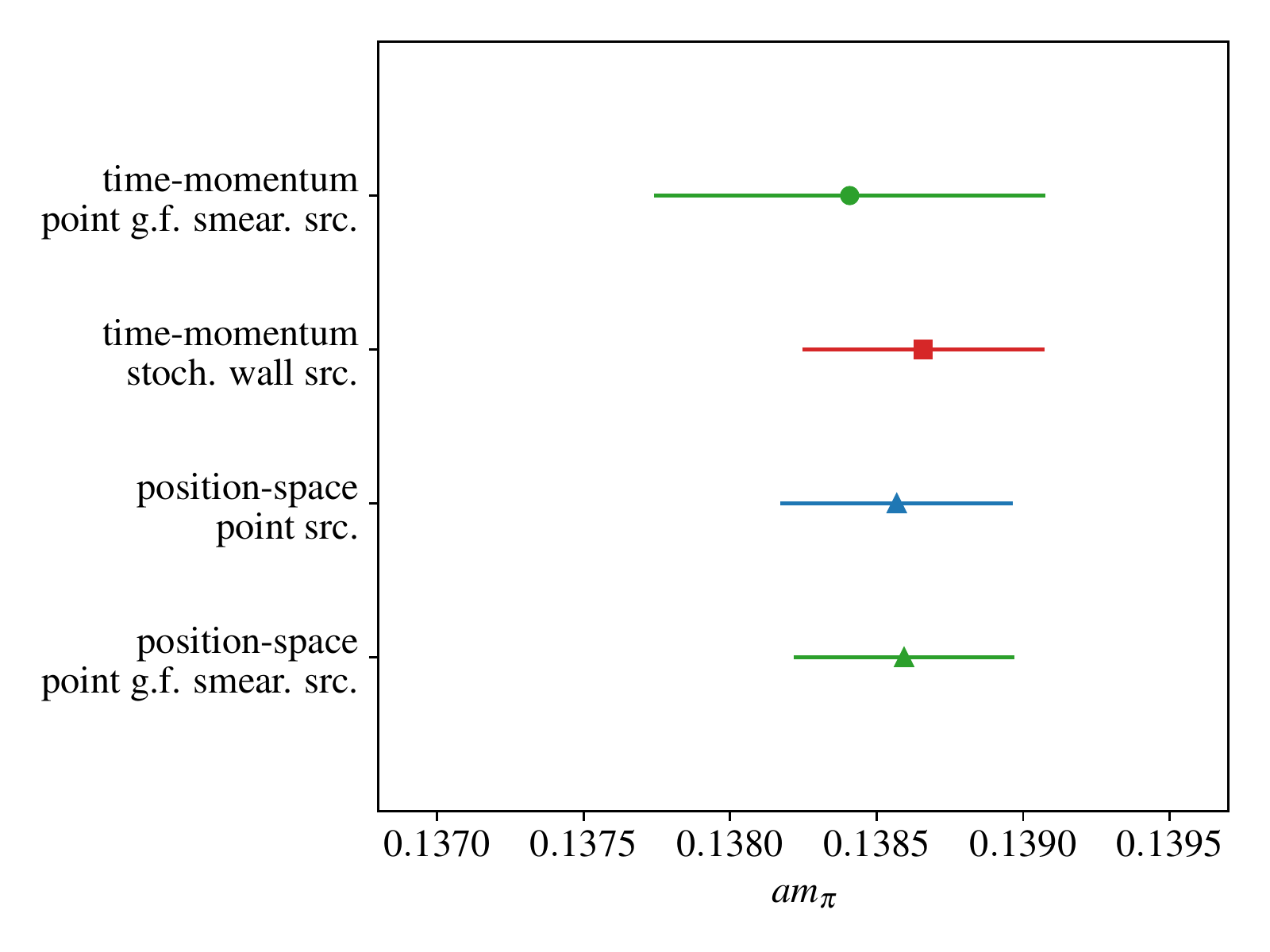}}
  \caption{%
    Left: effective mass corresponding to the radial pion correlator $\mathring{C}_{PP}(r)$.
    In lilac, the points obtained applying the infinite-volume formula for the long-distance behaviour in Eq.~\eqref{eq:CPPdef}.
    In blue, the points obtained modifying the long-distance behavior to account for finite-volume effects, Eq.~\eqref{eq:CPP_FV}.
    In green, the $r$-independent effective mass corresponding to the best \enquote{one-state} fit to the correlator.
    In red, the effective mass corresponding to the best \enquote{two-state} fit to the correlator as described in the main text.
    Right: comparison of the value of $m_\pi$ obtained from the \enquote{one-state} fit for various choices of correlators and smearing.
  }\label{fig:pion}
\end{figure}

In the left panel of Figure~\ref{fig:pion}, we plot the effective mass corresponding to the radial pion correlator $\mathring{C}_{PP}(r)$ with gradient-flow-smeared sources and sinks.
The points obtained by using Eq.~\eqref{eq:CPPdef} to define an effective mass according to Eq.~\eqref{eq:effective_mass_def} deviate significantly from the expected flat behaviour at large values of $r$.
We attribute this effect to the limited spacetime extent of the lattice, and we address this issue with a modification of the expected behaviour of the position-space correlator, performing a sum over all the images
\begin{equation}
  C_{PP}^{L,T}(x) = \sum_{n\in\mathbb{Z}^4} C(x+\mathbb{L}\cdot n) , \qquad \text{with $\mathbb{L}=\{T,L,L,L\}$} .
\end{equation}
Inserting the long-distance behaviour from Eq.~\eqref{eq:CPPdef} and performing the average over the $4d$ spheres, we obtain the radial correlator
\begin{multline}
\label{eq:CPP_FV}
  \mathring{C}_{PP}^{L,T}(r) = \sum_{n\in\mathbb{Z}^4} \frac{1}{\mathrm{r}_4(r^2)} \sum_{\abs{x}=r} C(x+\mathbb{L}\cdot n) \\
  \to \frac{\abs{c_P}^2}{4\pi^2} \left[ \frac{m_\pi}{r} K_1(m_\pi r) + 6 \int \frac{\dd{\Omega_3}}{2\cpi^2} \frac{m_\pi K_1(m_\pi\abs{x+L\cdot\hat{z}})}{\abs{x+L\cdot\hat{z}}} + \order{\e{-\sqrt{2} m_\pi L}, \e{-m_\pi T}} \right] ,
\end{multline}
where in the last line we included only the leading images from the finite $L$ extent.
This is analogous in spirit to the $\cosh$ effective mass used in the standard time-momentum representation.
Extracting the effective mass from the lattice correlator using Eq.~\eqref{eq:CPP_FV} results in a very flat behaviour at large values of $r$, as shown in the left panel of Figure~\ref{fig:pion}.

We also perform fits to the radial correlator directly, either using Eq.~\eqref{eq:CPP_FV} with the $\abs{c_P}^2$ and $m_\pi$ parameters as a fit ansatz, or adding to it an \enquote{excited state} $a_1 \frac{m_1}{r} K_1(m_1 r)$ with two extra parameters $a_1$, $m_1>m_\pi$.
With appropriately chosen fit ranges, the ground state parameters agree between the two fits, and the fitted $m_\pi$ agrees with the plateau average of the effective mass.
Moreover, the second fit is able to describe the correlator to much shorter radial distance.
We illustrate this by plotting the effective mass corresponding to the two fitted correlators in the left panel of Figure~\ref{fig:pion}.

In the right panel of Figure~\ref{fig:pion}, we compare the values of $m_\pi$ obtained from the \enquote{one-state} fit of both the time-momentum and the position-space correlators on the same point sources, both with and without smearing, and the time-momentum correlator on stochastic wall sources.
All the values obtained are compatible.
As expected for the meson correlator, we do not see any significant difference in the achieved precision between smeared point sources with either $3d$-fermion or gradient-flow smearing, and the non-smeared ones.
However, we see an increase in the precision of the point-sources determinations when changing from the time-momentum correlator to the position-space correlator one.
With the latter choice, the value of the pion mass is
\begin{equation}
\label{eq:pion_result}
  m_\pi = \num{0.1386(4)}/a \approx \SI{287.9(8)}{\MeV} ,
\end{equation}
and has a similar error to the value obtained, at the same computational cost, from the time-momentum correlator with stochastic wall sources.

\subsection{Results for the nucleon mass}

\begin{figure}[t]
  \centering
  \resizebox{\columnwidth}{!}{\includegraphics{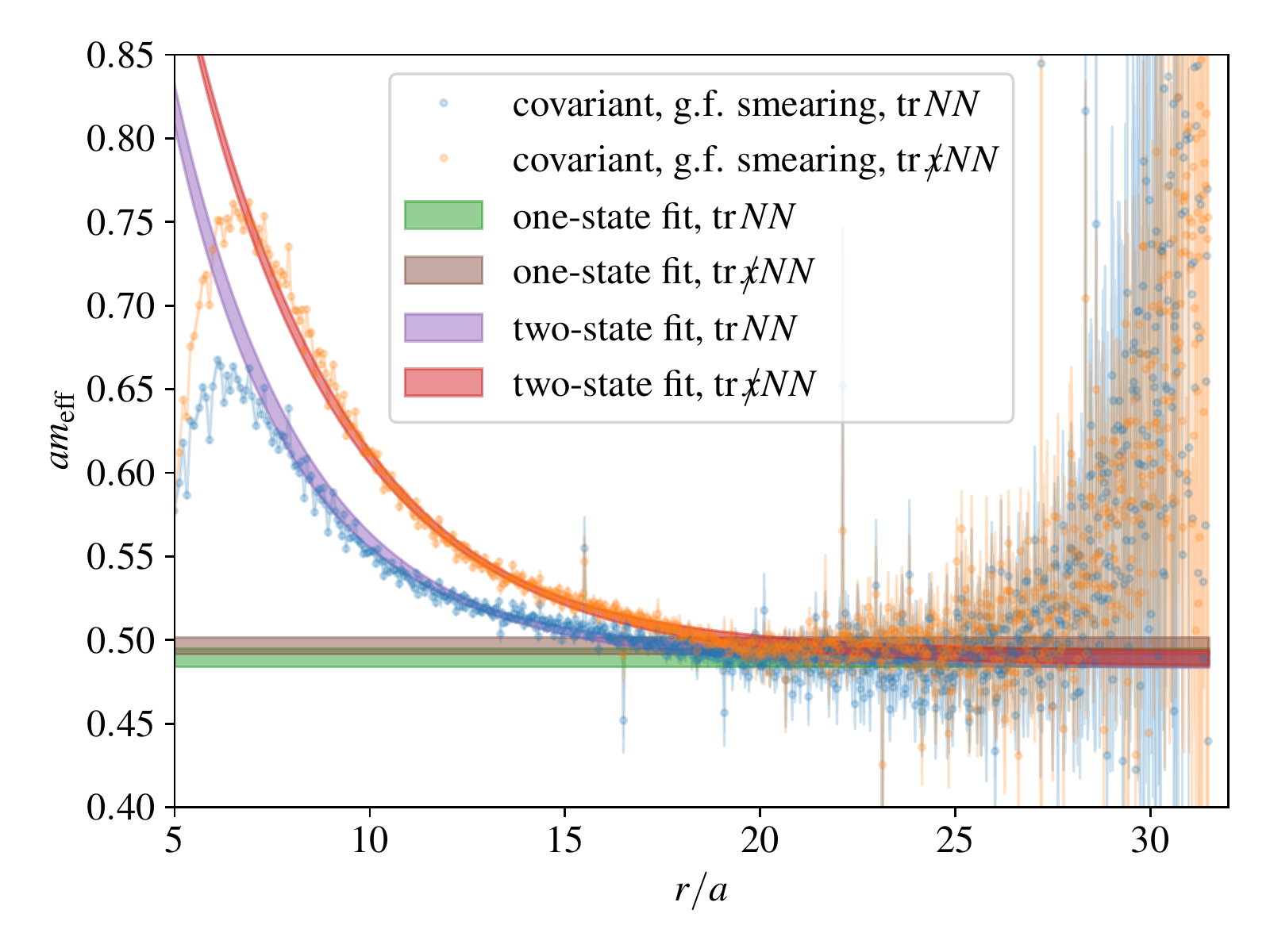}\includegraphics{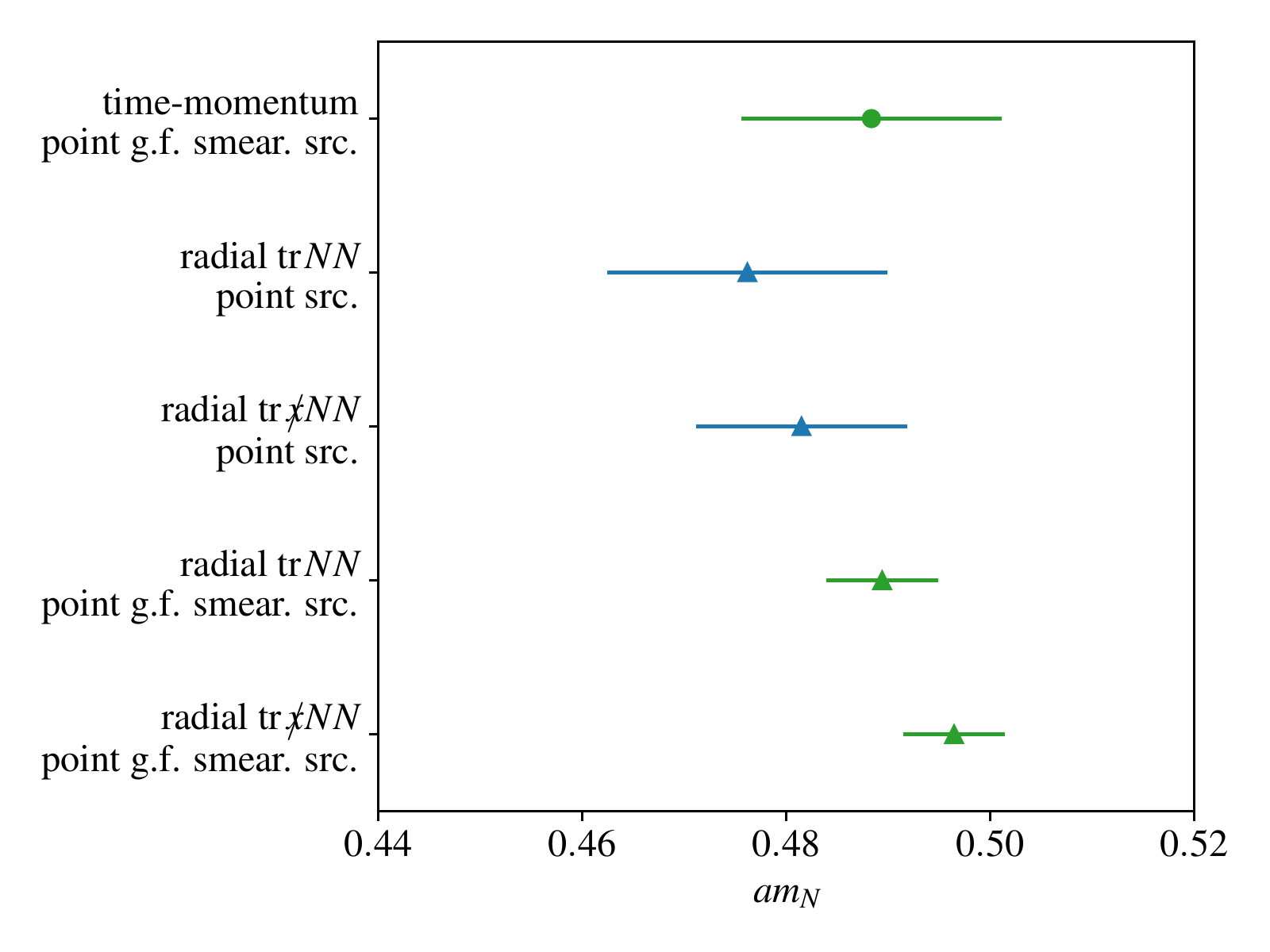}}
  \caption{%
    Left: effective mass corresponding to the radial nucleon correlator $\mathring{C}_{NN}(r)$.
    In blue, the points obtained applying the long-distance behaviour to the $\tr\mathring{C}_{NN}$ contraction in Eq.~\eqref{eq:trCNNdef}.
    In orange, the points obtained applying the long-distance behaviour to the $\tr\slashed{x}\mathring{C}_{NN}$ contraction in Eq.~\eqref{eq:trxCNNdef}.
    In green and brown, the $r$-independent effective mass corresponding to the best \enquote{one-state} fit to the $\tr\mathring{C}_{NN}$ and $\tr\slashed{x}\mathring{C}_{NN}$ correlators respectively.
    In violet and red, the effective mass corresponding to the best \enquote{two-state} fit to the $\tr\mathring{C}_{NN}$ and $\tr\slashed{x}\mathring{C}_{NN}$ correlators respectively as described in the main text.
    Right: comparison of the value of $m_N$ obtained from the \enquote{one-state} fit for various choices of correlators and smearing.
  }\label{fig:nucleon}
\end{figure}

The left panel of Figure~\ref{fig:nucleon} shows a similar analysis for the radial nucleon correlator $\mathring{C}_{NN}(r)$.
As in the case of the pion, the effective mass obtained by applying the long-distance behaviour in Eqs.~\eqref{eq:trCNNdef} and~\eqref{eq:trxCNNdef} shows a significant deviation from the flat behaviour that we attribute to finite volume effects.
In contrast to the case of the pion, the effect on the effective mass is positive. This corresponds to a correlator approaching zero with a faster rate than expected.
The main contribution to this effect does not come from mirror images, that fall off much faster than in the case of the pion, but from the propagation of intermediate $N\pi$ states.
Work to understand and correct for this effect is in progress, but for the purpose of this study we restrict fits to the correlator to $r_{\mathrm{max}}=24a$.
We also note that this effect is expected to become irrelevant on master-fields that have a much larger volume.

We perform fits to both contractions of the correlator using the functional form in either Eq.~\eqref{eq:trCNNdef} or Eq.~\eqref{eq:trxCNNdef} to extract $\abs{c_N}^2$ and $m_N$, and we compare them with fits including an additional excited state.
In this case, we find that $\mathring{C}_{NN}(r)\to\mathring{C}_{NN}(r)\cdot[1 + a_1 \frac{m_\pi}{r} K_1(m_\pi r)]$, with only one additional free parameter $a_1$ (with $m_\pi$ fixed from the fit to the pion correlator), adequately describes the behaviour of the correlator over a longer range of values of $r$.

The right panel of Figure~\ref{fig:nucleon} compares the results of the \enquote{one-state} fit for $m_N$ for different choices of smearing and contractions, both position-space and time-momentum, on the same point sources.
This time we observe that the smeared quark fields give more precise results, with gradient-flow smeared sources and sinks leading to a determination of $m_N$ that is about twice as precise as the unsmeared ones.
This is also true for the time-momentum determination, where we see that gradient-flow smearing performs better than our choices of $3d$-fermion smearing parameters, not shown in the figure.
More importantly, the mass extracted from a fit to the position space correlator is also twice as precise as the one from the time-momentum approach with the same smearing.
We observe a small difference between the two contractions in position space, with $\tr\slashed{x}\mathring{C}_{NN}$ corresponding to a systematically larger $m_N$.
We believe that these are discretization effects, which are in principle different between the two contractions.
From the $\tr\mathring{C}_{NN}$ contraction using gradient-flow smearing, we obtain the value
\begin{equation}
  m_N = \num{0.489(5)}/a \approx \SI{1015(10)}{\MeV} .
\end{equation}

\section{Conclusions and outlook}

The concept of stochastic locality, which is at the base of the master-field approach to lattice QCD, fits naturally with correlators defined in position space, in contrast with momentum-projected ones.
For these reasons, in these proceedings we have given an overview of practical methods to extract simple hadronic observables from position-space correlation functions of mesons and baryons.

Initially applying these methods to traditional ensembles with a moderately large volume, our numerical tests show a statistical accuracy competitive with standard time-momentum techniques: this is a promising result in view of the application to master-fields, addressing some of the challenges in making use of master-field translation averages for physics applications.
The main systematic effect comes from the finite volume of the lattice considered in the numerical tests.
We are able to fully correct for this systematics in the case of the pion, but not for the nucleon, which further strengthens the association of position-space techniques with master-fields in which volume effects are mostly negligible.
Moreover, while good results are obtained assuming rotational symmetry of the correlators, taking into account the symmetry breaking due to discretization effects remains an interesting topic for further studies.

With proper master-fields with volumes up to $(\SI{18}{\femto\metre})^4$ and $m_\pi L\approx\num{25}$ now available~\cite{Fritzsch:2021lattice}, we plan to combine the procedures discussed here with an error estimate fully based on translation averages.
A crucial step is to reduce the computational effort in hadronic observables from $\sim V^2$ to $\sim V$.
There are a number of strategies to achieve this that fit naturally with the choice of the position-space correlator, see \emph{e.g.}\ Ref.~\cite{Luscher:2017cjh}.
Similar ideas applied to the time-momentum correlator are also worth of investigation, such as the possibility of approximating the momentum projection by truncating the sum over $\vec{x}$ to make the correlator more local in space.
The hadronic observables that can profit from these new technological advances go beyond the simple ones considered here, and include cases in which the very large volume plays a crucial rôle, see Ref.~\cite{Bruno:2021lattice,Bulava:2021lattice}.

\section*{Acknowledgements}

We thank M.\ Lüscher for valuable discussions and comments on the proceedings draft.
This project has received funding from the European Union’s Horizon 2020 research and innovation programme under the Marie Skłodowska-Curie grant agreement No. 843134.
The research of MB is funded through the MIUR program for young researchers \enquote{Rita Levi Montalcini}.
The work of MTH is supported, in part, by UK Research and Innovation Future Leader Fellowship MR/T019956/1.

Many simulations were performed on a dedicated HPC cluster at CERN.
Part of the gauge fields were further generated on the HPC resource Frontera of the Texas Advanced Computing Center.
We gratefully acknowledge the computer resources and the technical support provided by these institutions.

\bibliographystyle{JHEP-journaltitle}
\bibliography{./biblio.bib}

\providecommand{\href}[2]{#2}\begingroup\raggedright\begin{thebibliography}{1}

\bibitem{Luscher:2017cjh}
M.~Lüscher, \emph{Stochastic locality and master-field simulations of very
  large lattices},
  \href{https://doi.org/10.1051/epjconf/201817501002}{\emph{EPJ Web Conf.}
  {\bfseries 175} (2018) 01002}
  [\href{https://arxiv.org/abs/1707.09758}{{\ttfamily 1707.09758}}].

\bibitem{Francis:2019muy}
A.~Francis, P.~Fritzsch, M.~Lüscher and A.~Rago, \emph{Master-field
  simulations of {$O(a)$}-improved lattice {QCD}: Algorithms, stability and
  exactness}, \href{https://doi.org/10.1016/j.cpc.2020.107355}{\emph{Comput.
  Phys. Commun.} {\bfseries 255} (2020) 107355}
  [\href{https://arxiv.org/abs/1911.04533}{{\ttfamily 1911.04533}}].

\bibitem{Fritzsch:2021lattice}
P.~Fritzsch, \emph{Master-field simulations of {QCD}},
  \href{https://doi.org/10.22323/1.396.0465}{\emph{PoS} {\bfseries LATTICE2021}
  (2021) 465}.

\bibitem{Giusti:2018cmp}
L.~Giusti and M.~Lüscher, \emph{Topological susceptibility at
  {$T>T_{\mathrm{c}}$} from master-field simulations of the {SU(3)} gauge
  theory}, \href{https://doi.org/10.1140/epjc/s10052-019-6706-7}{\emph{Eur.
  Phys. J. C} {\bfseries 79} (2019) 207}
  [\href{https://arxiv.org/abs/1812.02062}{{\ttfamily 1812.02062}}].

\bibitem{Luscher:2012av}
M.~Lüscher and S.~Schaefer, \emph{Lattice {QCD} with open boundary conditions
  and twisted-mass reweighting},
  \href{https://doi.org/10.1016/j.cpc.2012.10.003}{\emph{Comput. Phys. Commun.}
  {\bfseries 184} (2013) 519}
  [\href{https://arxiv.org/abs/1206.2809}{{\ttfamily 1206.2809}}].

\bibitem{Papinutto:2018ajw}
M.~Papinutto, F.~Scardino and S.~Schaefer, \emph{New extended interpolating
  fields built from three-dimensional fermions},
  \href{https://doi.org/10.1103/PhysRevD.98.094506}{\emph{Phys. Rev. D}
  {\bfseries 98} (2018) 094506}
  [\href{https://arxiv.org/abs/1807.08714}{{\ttfamily 1807.08714}}].

\bibitem{Luscher:2013cpa}
M.~Lüscher, \emph{Chiral symmetry and the yang-mills gradient flow},
  \href{https://doi.org/10.1007/jhep04(2013)123}{\emph{JHEP} {\bfseries 1304}
  (2013) 123} [\href{https://arxiv.org/abs/1302.5246}{{\ttfamily 1302.5246}}].

\bibitem{Bruno:2021lattice}
M.~Bruno, \emph{Variations on the {Maiani}-{Testa} approach and the inverse
  problem}, \href{https://doi.org/10.22323/1.396.0405}{\emph{PoS} {\bfseries
  LATTICE2021} (2021) 405}.

\bibitem{Bulava:2021lattice}
J.~Bulava, \emph{Spectral reconstruction of an inclusive rate in the
  two-dimensional {$\mathrm{O}(3)$} model},
  \href{https://doi.org/10.22323/1.396.0304}{\emph{PoS} {\bfseries LATTICE2021}
  (2021) 304}.

\end{thebibliography}\endgroup

\end{document}